\documentclass[twocolumn,secnumarabic,amssymb, nobibnotes, aps, prl, superscriptaddress, nobalancelastpage,longbibliography]{revtex4-2}

 \usepackage{graphicx}
\usepackage{booktabs} 
\usepackage{multirow}
\usepackage{bm}
\usepackage{amsmath} 
\usepackage{braket} 
\usepackage{epsfig}
\usepackage{tensor}
\usepackage{CJKutf8}
\usepackage[version=4]{mhchem}
\usepackage{titlesec}
\usepackage{ifthen}
\usepackage{sidecap}
\usepackage{listings}
\usepackage{array}
\usepackage{floatrow}
\usepackage{enumitem}
\usepackage[para,online,flushleft]{threeparttablex}
\usepackage{diagbox}
\usepackage{float}
\floatstyle{plaintop}
\restylefloat{table}

\usepackage{hyperref}
\usepackage{xr}
\hypersetup{breaklinks=true,colorlinks=true,linkcolor=blue,citecolor=blue,filecolor=magenta,urlcolor=cyan}

\usepackage[all]{hypcap}

\usepackage{xcolor}
\definecolor{pastelgray}{rgb}{0.81, 0.81, 0.77}
\definecolor{beaublue}{rgb}{0.9, 0.9, 0.93}

\usepackage{orcidlink}
\usepackage{tikz,xcolor,hyperref}

\definecolor{lime}{HTML}{A6CE39}
\DeclareRobustCommand{\orcidicon}{
	\begin{tikzpicture}
	\draw[lime, fill=lime] (0,0) 
	circle [radius=0.16] 
	node[white] {{\fontfamily{qag}\selectfont \tiny ID}};
	\draw[white, fill=white] (-0.0625,0.095) 
	circle [radius=0.007];
	\end{tikzpicture}
	\hspace{-2mm}
}

\foreach \x in {A, ..., Z}{%
	\expandafter\xdef\csname orcid\x\endcsname{\noexpand\href{https://orcid.org/\csname orcidauthor\x\endcsname}{\noexpand\orcidicon}}
}

\renewcommand{\vec}[1]{\mbox{\boldmath $#1$}}

\makeatletter
\def\@bibdataout@aps{%
\immediate\write\@bibdataout{%
@CONTROL{%
apsrev41Control%
\longbibliography@sw{%
    ,author="08",editor="1",pages="1",title="0",year="1"%
    }{%
    ,author="08",editor="1",pages="1",title="",year="1"%
    }%
  }%
}%
\if@filesw \immediate \write \@auxout {\string \citation {apsrev41Control}}\fi
}
\makeatother

\renewcommand{\vec}[1]{\mbox{\boldmath $#1$}}

\newcolumntype{Y}{>{\centering\arraybackslash}X}

\begin{document}

\begin{CJK*}{UTF8}{gbsn}

\title{Nuclear Radii of Proton-Unbound Systems}

\author{Y. R. Lin (林雅茹)}
\affiliation{Key Laboratory of Nuclear Physics and Ion-beam Application (MOE), Institute of Modern Physics, Fudan University, Shanghai 200433, China}
\affiliation{Shanghai Research Center for Theoretical Nuclear Physics,
NSFC and Fudan University, Shanghai 200438, China}

\author{S. M. Wang (王思敏)\,\orcidlink{0000-0002-8902-6842}}\email{Email: wangsimin@fudan.edu.cn}
\affiliation{Key Laboratory of Nuclear Physics and Ion-beam Application (MOE), Institute of Modern Physics, Fudan University, Shanghai 200433, China}
\affiliation{Shanghai Research Center for Theoretical Nuclear Physics,
NSFC and Fudan University, Shanghai 200438, China}

\author{W. Nazarewicz\,\orcidlink{0000-0002-8084-7425}}\email{Email: witek@frib.msu.edu}
\affiliation{Facility for Rare Isotope Beams, Michigan State University, East Lansing, Michigan 48824, USA}
\affiliation{Department of Physics and Astronomy, Michigan State University, East Lansing, Michigan 48824, USA}

\begin{abstract}
Nuclear radius is a fundamental structural observable that informs many properties of atomic nuclei and nuclear matter. 
Experimental studies of radii in drip line nuclei are in the forefront of research with radioactive ion beams. Of particular interest are charge radii of proton-unbound nuclei that will soon be approached in laser spectroscopy.
In this Letter, using the complex-energy approach and direct time propagation, we investigate the radius of  the  proton resonance whose size is ill defined in the  standard stationary quantum-mechanical description. 
An early-time plateau is identified during which the 
radius of the Gamow resonance coincides with the real-energy radius accessible experimentally. We demonstrate a 
nonmonotonic dependence of the complex radius on decay energy
and  a local increase of the charge radius across the threshold
(a halolike enhancement).

\end{abstract}

\date{\today}

\maketitle

\end{CJK*}


\textit{\label{sec:level1}Introduction}---The limits of the nuclear landscape are formally marked by particle drip lines, beyond which the nucleonic decays are  possible because of positive decay $Q$ values. Moving away from drip lines by adding protons or neutrons, one encounters a zone of metastable states (long-lived resonances) that are amenable to experimental investigations. This zone is fairly extended on the proton-rich side because of the presence of the Coulomb barrier. Further away from the drip lines, one enters the ephemeral zone of very short-lived nuclear states where the very notion of nuclear existence may become questionable 
\cite{Thoennessen2004,Nazarewicz2025}.

Proton emitters are narrow resonances beyond the proton drip line\cite{Goldansky1960,Blank2008,Blank2023,Pfutzner2012,Pfutzner2023,Pfützner2023a}. Since their lifetimes primarily depend on $Q$ values and angular momentum, studies of proton emitters provide unique  information on nuclear structure and reactions  in the presence of the low-lying proton continuum. Energetically,  proton-emitting nuclei
are clustered in a  rather  narrow window of $Q$ values \cite{Olsen2013,Neufcourt2020}. For the large  $Q_{1p/2p}$ values, decay lifetimes are  going to be too fast to be observed ($T_{1p/2p}<100$\,ns). On the other hand, if $Q_{1p/2p}$ values are  too low  ($T_{1p/2p}>100$\,ms), proton-decay rates  cannot compete with  other decay modes, 
such as $\beta^+$, electron capture, or $\alpha$ decays.

Matter radii of unstable nuclei can be studied using hadronic probes  by measuring reaction and interaction cross sections \cite{Kanungo2023,Alkhazov2011}. The  
proton and neutron radii extracted in this way, are prone to appreciable uncertainties because of reaction modeling. The precise information on  sizes of unstable nuclei comes, therefore,  from studies of nuclear charge radii.
Indeed, due to experimental progress in laser spectroscopy, nuclear charge radii can now be measured along long isotopic chains of short-lived isotopes \cite{Otten1989,Nörtershäuser2023}.
The future experimental programs in this area intend to investigate charge radii
of nuclei close to and beyond the proton drip line \cite{Cheal2022,Lynch2024}.
This exciting perspective raises several theoretical challenges.

For nuclei that are formally bound, but close to drip line, the low-energy scattering continuum can impact nuclear radii.  Here, excellent examples are nuclear halos \cite{Jensen2004} and  radial properties of proton-rich 
nuclei  such as $^{36}$Ca \cite{Miller2019,Reinhard2022,Xu2025}. For unbound nuclear states, the notion of the root-mean-square (rms) radius becomes problematic as resonances  are not stationary. For extremely narrow states,
it is customary to  use the bound-state approximation, in which the scattering tail of the wave function is neglected. For states with shorter lifetimes, however, the real-energy stationary approach cannot be used as the scattering tail of the one-body density would formally result in 
an infinite rms radius. The two alternative methods are (i) the time-dependent formalism based on the explicit solution of the time-dependent Schr\"odinger equation (TDSE) or (ii) the stationary complex-energy  resonant-state approach in which the outgoing boundary condition is imposed. While the former method is not effective for narrow resonances, the latter can be used for both narrow and wide resonant states. While the complex-energy framework has been successfully used to describe radii of halo nuclei \cite{Myo2010,Papadimitriou2011,Kruppa2014,Myo2014}, many  questions related to the interpretation of complex expectation values 
in resonant states remain \cite{Berggren1996,Burgers1996,Bollini19961,Bollini19962,Homma1997,Civitarese1999,Sekihara2013,Papadimitriou2016,Myo2023}.

As stated above, due to their long lifetimes, proton emitters are splendid examples of complex-energy resonant states 
\cite{Ferreira1997,Maglione1999,Rykaczewski1999,Kruppa2000,Barmore2000,Wang2017,Wang2018}. Broad proton resonances can also be described in terms of the TDSE \cite{Talou1998,Talou1999,Oishi2014,Oishi2023,Oishi2025,Wang2021,Wang2023}. In this study, we use both methods to investigate and interpret  complex rms radii of proton-unbound  states.

\textit{Method}\label{the model}
---This study focuses on single-proton emitters, in which the  valence proton  is coupled to the daughter nucleus (core) \cite{Buck1992,Aberg1997}. The effective radial nucleon-core potential, $V(r)$, includes the nuclear interaction in the form of a Woods-Saxon (WS) potential with the  spin-orbit term, the Coulomb interaction, and the centrifugal potential. 
 The solutions of the Schr\"odinger equation with the potential $V$ that have purely outgoing asymptotic are resonant (Gamow) states.
 The resonant eigenfunction $\Phi^{J\ell M}(k,\boldsymbol{r})=\psi_{J\ell}(k,r)Y_{J\ell M}(\Omega)$ has angular quantum numbers $(J,\ell,M)$ and corresponds to momentum $k$ with energy $E=\hbar^2 k^2/2\mu$. To ensure the outgoing asymptotic behavior, the radial wave function $u_{J\ell}(k,r) = r\psi_{J\ell}(k,r)$ is expanded in the Berggren basis \cite{Berggren1968,Michel2009}. The corresponding eigenenergy is complex, $\tilde{E} = Q_p - i\Gamma/2$, where $Q_p$ is the decay energy and $\Gamma$ is the decay width related to the half-life $T_{1/2} = \hbar \ln 2 / \Gamma$.
 
 Unlike in the standard Hilbert-space framework, resonant  wave functions 
 belong to the  rigged Hilbert space (RHS) \cite{Antoine2009,Michel2021}; they are
 not $L^2$ square integrable. In RHS, both bound and resonant states are normalized according to a  biorthogonal inner product \cite{Moiseyev1978}. For resonant states, the radial wave function $u_{J\ell}(k,r)$ asymptotically approaches the outgoing Coulomb wave function $H_{\ell,\eta}^+(kr)$, scaled by an asymptotic normalization coefficient (ANC) $a_{J\ell}(k)$ where $\eta $ is the Sommerfeld parameter. Since $H_{\ell,\eta}^+(kr)$ exhibits oscillatory behavior at large $r$, the contribution from the asymptotic region to the normalization integral largely cancels out.
 
To avoid the divergence of the rms radius of resonant states, a complex radius $\tilde{r}_{\rm rms}$ can be introduced, defined as
\begin{equation}\label{complex_radius} 
\tilde{r}_{\rm rms}^2 
\equiv \langle \tilde{\psi} | r^2 | \psi \rangle 
= \int_{0}^{\infty} (r \psi)^2 r^2  \, dr.
\end{equation} 
To evaluate this integral, the exterior complex scaling method is employed~\cite{Gyarmati1971}
by introducing the rotated complex contour ${\cal C}$ corresponding to the complex-scaled radius
 $\tilde{r}$: 
\begin{equation}
\label{coordinate transformation}
\tilde{r} = 
\begin{cases}
r, & \text{for } r \leq R_0; \\
R_0 + (r - R_0)\, e^{i\theta}, & \text{for } r > R_0,
\end{cases}
\end{equation}
where \( R_0 \) is the end point at which the complex rotation is applied and $\theta$ is the rotation angle. Under this transformation, Eq.\,(\ref{complex_radius}) becomes
\begin{equation}\label{complex_radius_integral}
\tilde{r}_{\rm rms}^2
= \int_{0}^{R_0} r^4\psi^2 \, dr 
- \int_{{\cal C}} \tilde{r}^2 a_{J\ell}^2\, ({H^+_{\ell,\eta}})^2\, d\tilde{r} .
\end{equation}
The convergence at large $|\tilde{r}|$ is ensured by choosing $\theta > \arctan[{{\Im}(\tilde{k})}/{{\Re}(\tilde{k})}]$, where $\Im$ and $\Re$ denote the imaginary and real parts, so that the asymptotic factor $e^{ik r}$ (valid for $r\gg R_0$) decays exponentially along the rotated complex contour $\mathcal{C}$. 

To relate the complex radius in the resonant state to the measured rms radius, we employ the time-dependent approach \cite{Wang2021} to track the evolution of the decaying nucleon.  Here, the initial state is modeled by means of  the two potential approach (TPA) \cite{Gurvitz1987,Gurvitz2004}, in which the closed initial potential $V_{\rm TPA}(r)$ is assumed to be  constant $V(r_{\rm TPA}) > Q_p$ for $r\ge r_{\rm TPA}$ [see Fig.\,S1 in Supplemental Material (SM) \cite{SM}  for illustration], where $r_{\rm TPA}$ is chosen inside the barrier (between the barrier radius and outer turning point)
\cite{Gurvitz2004}. As shown below, the evolution during the moderate time period is governed by the resonant dynamics of the state and is essentially insensitive to the choice of $r_{\rm TPA}$.

\nocite{Fonda1978,Deutsch2018,Gorin2006,Volya2020,Gurvitz2004}

Our analysis is carried out for the rms radius of the $d_{5/2}$ resonant state in the proton emitter $^{15}$F($^{14}$O+$p$), a medium-mass nucleus with a moderate decay width ~\cite{Mukha2010,Girard-Alcindor2022}, which constitutes an excellent laboratory for various open quantum system phenomena~\cite{Okolowicz2009,Okolowicz2011,Girard-Alcindor2022,Oishi2025}.


\textit{Hamiltonian and parameters}\label{the Hamiltonian}---Unless stated otherwise, the exterior complex scaling is carried out with  $R_{0} = 20$\,fm and the rotation angle $\theta = \pi/4$. The WS depth $V_0$ is adjusted to set the resonance energy, while all other parameters follow the ``universal'' WS    parametrization~\cite{Dudek1981}. The Berggren basis is built on the contour  $k = 0 \rightarrow 0.3-0.2i\rightarrow 0.5-0.15i\rightarrow 0.6\rightarrow0.8\rightarrow 1.2\rightarrow 2\rightarrow 4\rightarrow 6$ fm$^{-1}$, with 80 points per segment (numerical uncertainty $<1\%$). For the real-time propagation, we project the contour onto the real axis to restore Hermiticity and conserve norm, expanding the initial wave function in a real-momentum basis within $R_{\rm cut}=50$ fm. Results are practically unchanged provided $R_{\rm cut}$ exceeds the spatial extent of the localized initial state and the analysis region (see SM videos~\cite{SM}). The matching radius $r_{\rm TPA}$ is chosen to match the radii from the complex-scaling and time-dependent analyses.


\begin{figure}[htb]
\includegraphics[width=0.9\linewidth]{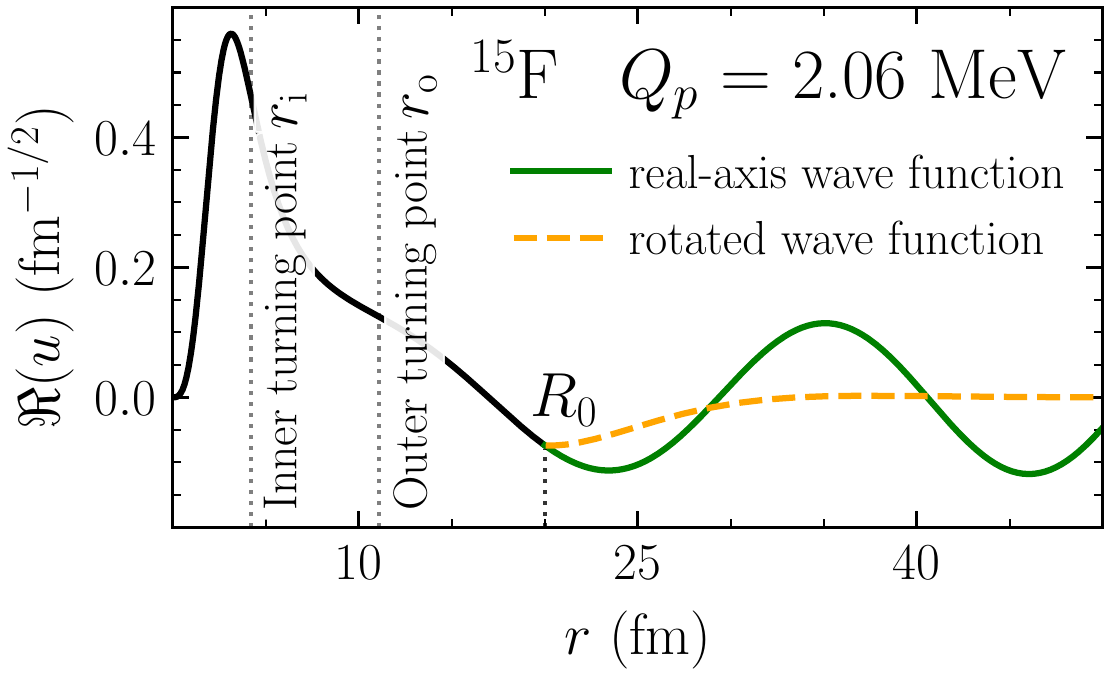}
\caption{
\label{fig:Complex_radius_scaling}
Real part of the $d_{5/2}$ radial wave function $u$ 
in $^{15}$F at $Q_p=2.06$\,MeV 
evaluated along the real axis $r$ (solid line) and along the complex-rotated coordinate $\tilde{r}$   (dashed line), see Eq.~(\ref{coordinate transformation}).
The  inner ($r_{\rm i}$) and outer  ($r_{\rm o}$) turning points are indicated by dotted lines.}
\end{figure}

\textit{Complex radius}---Figure~\ref{fig:Complex_radius_scaling} shows the wave function of the resonant state in $^{15}$F at $Q_p=2.06$\,MeV along the real-$r$ axis and along the rotated contour $\cal C$. The wave function is localized in the nuclear interior, and it decays exponentially inside the barrier. Outside the outer turning point $r_{\rm o}$, the outgoing boundary condition induces the expected oscillatory behavior of the resonant wave function $u(r)$.
The rotation radius $R_0$ defining the contour $\cal C$ is chosen beyond $r=r_{\rm o}$. With this choice, the rotated wave function $u(\tilde{r})$ is exponentially damped and rapidly vanishes along the contour.
As shown in Table~S1 of SM  \cite{SM}, the computed complex rms radii are practically independent of the choice of $\cal C$.

\begin{figure}
\centering
\includegraphics[width=1.0\linewidth]{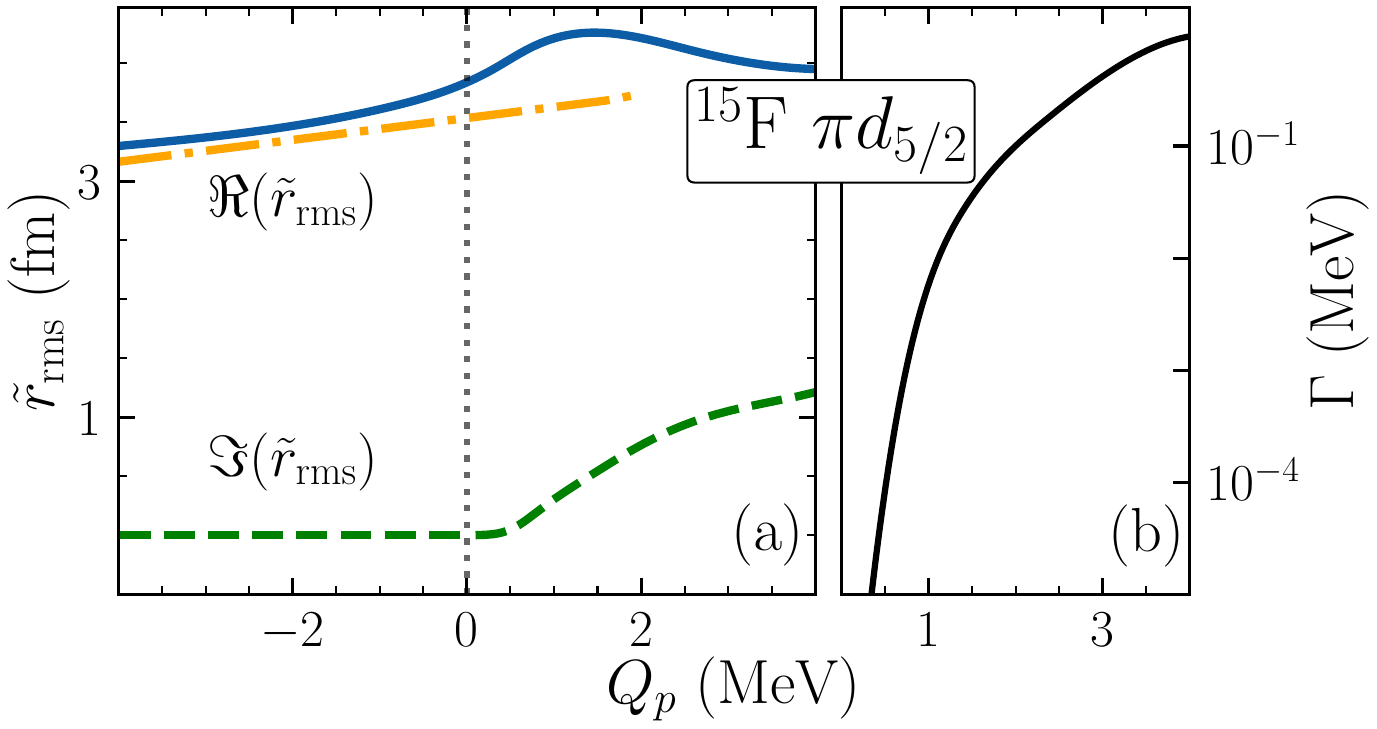}
\caption{
(a) Complex-rms radius $\tilde{r}_{\rm rms}$ and (b)  decay width $\Gamma$ as a function of decay energy $Q_p$  calculated for the proton resonant $d_{5/2}$ state in $^{15}\mathrm{F}$. The real and imaginary parts of $\tilde{r}_{\rm rms}$ are shown by the  solid and  dashed lines, respectively. The  dash-dotted line shows the rms radius in the HO basis, i.e., without continuum coupling. The proton threshold is marked by the  dotted line.
}
\label{Radius-Energy}
\end{figure}

For bound states, the complex radius $\tilde{r}_{\rm rms}$ reduces to the usual rms radius $r_{\rm rms}
$. Once the state crosses the decay threshold, it becomes a resonance characterized by a complex energy $\tilde{E}$, where the real part $Q_p$ corresponds to the average value of energy and the imaginary part $\Gamma$ is related to the energy uncertainty~\cite{Berggren1968, Berggren1996}. The same interpretation applies to any complex quantity commuting with the Hamiltonian~\cite{Berggren1968, Berggren1996}, and can be extended to self-adjoint operators---such as the radius operator---within the framework of tempered ultradistributions and Gel'fand triplets by considering the leading-order terms in $\Gamma$~\cite{Bollini19961,Bollini19962,Civitarese1999}.

As shown in Fig.~\ref{Radius-Energy}(a), the imaginary part of the complex radius increases with the decay energy, correlating strongly with the growth of the decay width~\cite{Gyarmati1972} in Fig.~\ref{Radius-Energy}(b). In contrast, the real part exhibits a nonmonotonic trend: it initially increases as the wave function becomes more diffuse in the asymptotic region during evolution from a bound to a resonant state, but beyond a certain point it decreases with increasing decay energy. As discussed below, 
this behavior reflects the interplay between the spatial extension and the finite lifetime of the resonant wave function.


To gauge the role of continuum coupling, we also compute the $d_{5/2}$ state in a localized harmonic-oscillator (HO) basis, using the lowest four radial-node configurations as a bound-state-like baseline that lacks explicit continuum effects. As shown in Fig.~\ref{Radius-Energy}(a), the Berggren–HO difference gradually increases from $Q_p\approx -1$\,MeV and becomes quite appreciable above the proton threshold, indicating a halolike enhancement.

\begin{figure}[htbp]
\includegraphics[width=1\linewidth]{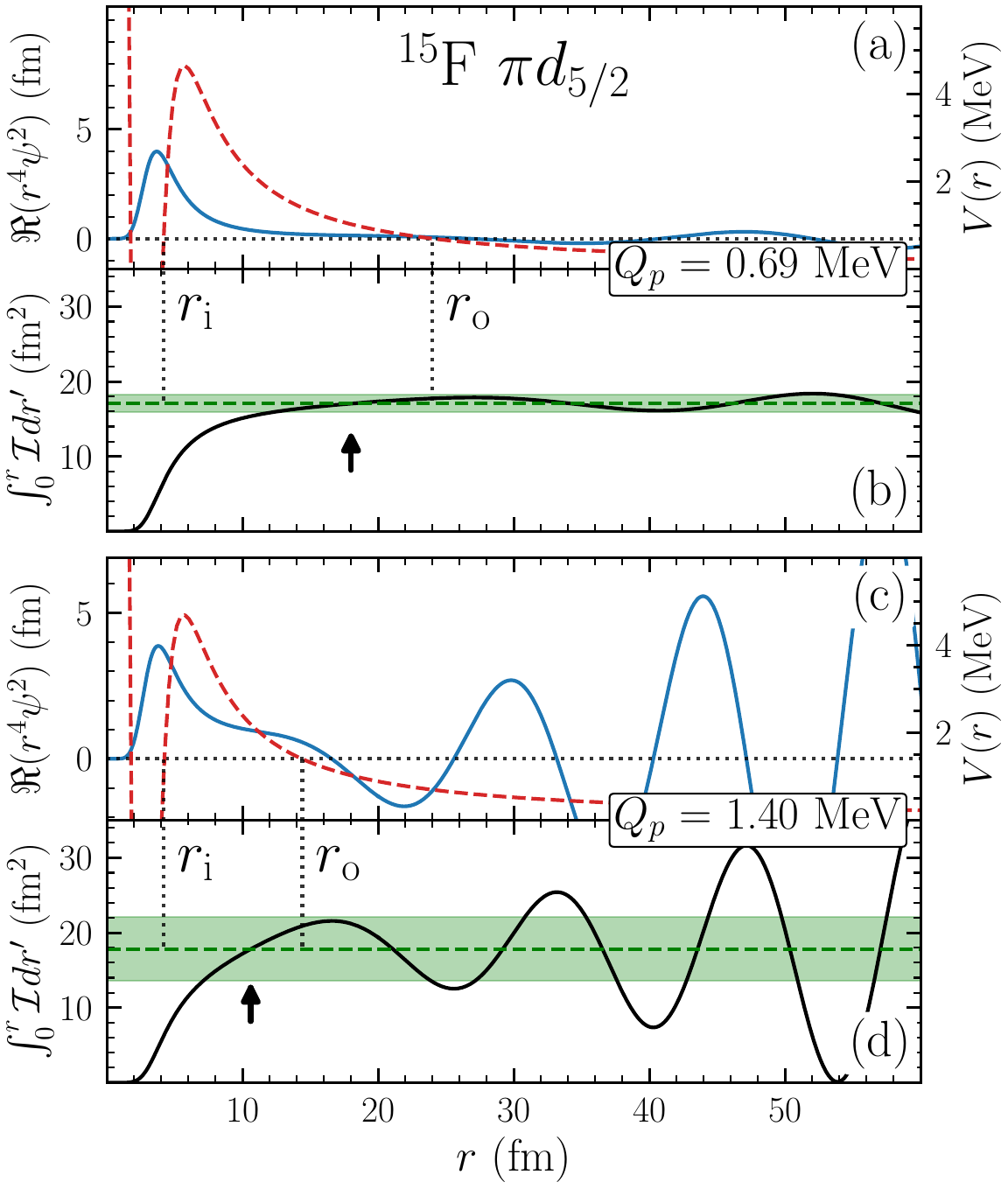}
\caption{\label{fig:int_15F_d}
Complex radius integrand $r^4 \psi^2$ (a), (c) and the corresponding cumulative integral $\int_0^r r^{\prime 4} \psi^2 \, d r^\prime$ (b), (d) for the proton resonant $d_{5/2}$ state in $^{15}$F at $Q_p=0.69$\,MeV (top) and 1.40\,MeV (bottom). All quantities are evaluated along the real axis without complex rotation. The  dashed line indicates the potential barrier. The  dash-dotted line and green shaded band mark the real and imaginary part of the complex radius $\tilde{r}_{\rm rms}^2$, respectively. The  inner  and outer turning points are indicated. The intersection point at which  the cumulative integral  becomes equal to  the complex rms radius in shown by an arrow.}
\end{figure}

To understand this nonmonotonic behavior of $\Re(\tilde{r}_{\rm rms})$ seen in Fig.~\ref{Radius-Energy}(a), in Fig.\,\ref{fig:int_15F_d} we examine the integrand $r^4 \psi^2$ defining  the complex rms radius, along with the cumulative integral $\int_0^r r^{\prime 4} \psi^2 \, d r^\prime$, both plotted along the real coordinate $r$. The wave function of the resonant state  exhibits oscillatory behavior with increasing amplitude, as expected for a Gamow state~\cite{Baz1969}. Since the contribution to the rms radius  from the asymptotic region is largely canceled out, it is  the internal part of the wave function that  determines the complex radius.


Importantly, the cumulative integral of $r^{4}\psi^{2}$ reaches $\tilde r_{\rm rms}^{2}$ near the outer turning point. With increasing decay energy, the barrier becomes lower and narrower, reducing the tunneling region and depleting the interior amplitude. At sufficiently high energies, this  can lead to a modest decrease of the rms radius. The detailed energy dependence is system dependent and merits further study.

\textit{Time dependent analysis}---The concept of complex observables has been broadly employed across atomic, hadronic, and nuclear physics~\cite{Berggren1996,Bollini19961,Civitarese1999,Sekihara2013,Papadimitriou2016,Myo2014,Myo2023}. Yet, a direct demonstration linking these complex quantities to measurable observables remains lacking. In reality, a resonant state decays within a finite lifetime, during which its spatial distribution evolves from an initially localized wave packet into outgoing decay fragments, accompanied by a change in its radius. We quantify this process by computing the time-dependent rms radius for various decay energies and comparing it with the complex rms radius. The initial time  $t=0$ is defined as the moment immediately following the formation, when the decaying proton is confined within the parent nucleus. This wave function  is subsequently propagated with the Hermitian Hamiltonian in the standard Hilbert space.

\begin{figure}[htbp]
\includegraphics[width=1\linewidth]{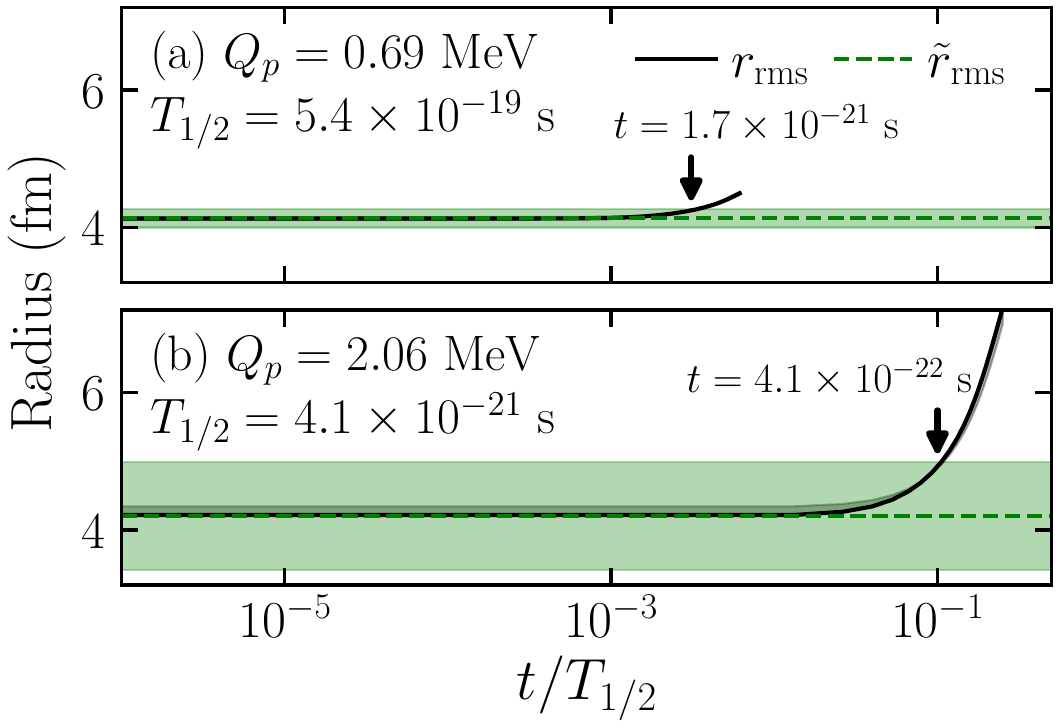}
\caption{\label{fig:t_r_cr_15F} Time evolution of the rms radius \(r_{\rm rms}\) of the $d_{5/2}$ resonance in \(^{15}\mathrm{F}\) (solid line) as a function of time $t$ (in units of  the half-live $T_{1/2}$) for (a) \(Q_p=0.69\) MeV and (b) \(Q_p=2.06\) MeV. The real  (dashed line) and imaginary 
(shaded band) parts of the complex radius \(\tilde{r}_{\rm rms}\) are shown for comparison. The uncertainty  in  \(r_{\rm rms}\) 
related to the variation of  \(r_{\rm TPA}\) by \(\pm 1\)\,fm is marked by a gray band.
 Arrows mark the time at which \(r_{\rm rms}
\) start departing    from \(\Re(\tilde{r}_{\rm rms})\) within the uncertainty  \(\Im(\tilde{r}_{\rm rms})\).}
\end{figure}


During the time evolution, the valence nucleon is emitted and the system's wave function becomes progressively more diffuse (see Figs. S2 and S3 in SM~\cite{SM}), leading to an eventual increase of $r_{\rm rms}$. As shown in Fig.~\ref{fig:t_r_cr_15F}, the growth rate correlates with the width $\Gamma$ (or $T_{1/2}$), while the survival probability shows  exponential decay over a broad time window (Fig.~S6 in the SM~\cite{SM}), confirming the accuracy of the propagation. Evaluating $r_{\rm rms}$ is more delicate because the wave function is evolved in momentum space (Berggren basis) but the radius is computed in coordinate space; reliable results require a well-converged tail within the analysis box. To control this effect, we therefore restrict our analysis to early times with total leakage $<0.01\%$. This limitation, however,  does not affect $|\langle\Phi(t)|\Phi(0)\rangle|^2$ (see Fig.\,S6 in SM \cite{SM}) because the localized initial state  suppresses the escaped flux.

Theoretically, the squared rms radius \(r_{\rm rms}^2\) follows an approximately quadratic behavior at the early stage of the decay process,
\begin{equation}
\label{r_rms_TD_main}
r_{\rm rms}^2(t)\simeq r_{\rm rms}^2(0)+b_{\rm TPA}\,t^2,
\end{equation}
with $b_{\rm TPA}=\langle \vec r\!\cdot\!\vec\nabla(V_{\rm TPA}-V)/m\rangle_{t=0}$ (see SM~\cite{SM}). Since $V_{\rm TPA}-V$ differs only in the asymptotic region --- where the quasi-bound initial wave function is negligible and the potential varies smoothly --- \(b_{\rm TPA}\) is small in magnitude, yielding an early-time plateau: for $t\lesssim10^{-3}T_{1/2}$, $r_{\rm rms}$ is essentially constant and coincides with the complex radius. This plateau reflects the generic nonexponential short-time regime with vanishing decay rate at $t=0$~\cite{Fonda1978,Deutsch2018,Gorin2006,Volya2020,Peshkin2014} and its connection to state preparation and quantum-Zeno physics~\cite{Ghirardi1979,Chiu1977, Kofman1996,Wilkinson1997}; thus, for sufficiently long-lived systems, a measurement at times $\ll T_{1/2}$ should be consistent with the complex radius. At later times, $r_{\rm rms}^2$ continues to grow slowly and roughly quadratically, consistent with a ``classical-motion'' form where $b_{\rm TPA}/r_{\rm rms}(0)$ acts as an effective acceleration (see SM~\cite{SM} for the derivation).  This motivates an operational ``escape time,'' defined as the time when $r_{\rm rms}$ first exceeds the intrinsic uncertainty band set by $\Im(\tilde r_{\rm rms})$ (arrows in Fig.~\ref{fig:t_r_cr_15F}); beyond it, the stationary-resonance picture breaks down and the dynamics crosses over from a localized nucleus to an effectively outgoing proton. Consistently, Fig.~\ref{fig:int_15F_d} and the SM videos~\cite{SM} show that while the evolution stays within this band, the radius is dominated by the nuclear interior.

To assess sensitivity to initial-state preparation, we vary $r_{\rm TPA}$ by $\pm1$ fm. The induced spread in $r_{\rm rms}$ (gray band in Fig.~\ref{fig:t_r_cr_15F}) is negligible, showing that the early-time plateau is essentially independent of initial conditions and thus provides an operational characterization of the resonance; at intermediate times, nonresonant components disperse rapidly, and the evolution becomes resonance dominated~\cite{Wang2021,Wang2023}. We also scan the physically relevant range of $r_{\rm TPA}$ from the barrier top to the outer turning point $r_{\rm o}$ (6--23 fm for $Q_p=0.69$ MeV; Fig.~\ref{fig:int_15F_d}). Although $b_{\rm TPA}$ depends only weakly on $r_{\rm TPA}$ and the initial $r_{\rm rms}$ increases smoothly with $r_{\rm TPA}$ (Fig.~S5 in the SM~\cite{SM}), the resulting values lie largely within the intrinsic uncertainty band $\Im(\tilde r_{\rm rms})$, centered at $\Re(\tilde r_{\rm rms})$. This supports interpreting $\tilde r_{\rm rms}$ as the mean configuration formed inside the barrier and justifies choosing $r_{\rm TPA}$ to match the complex-scaling and time-dependent radii; interestingly, the condition $r_{\rm rms}=\Re(\tilde r_{\rm rms})$ occurs near the minimum of the $r_{\rm rms}$ growth rate versus $r_{\rm TPA}$.

\begin{figure}[htbp]
\centering
\includegraphics[width=1.0\linewidth]{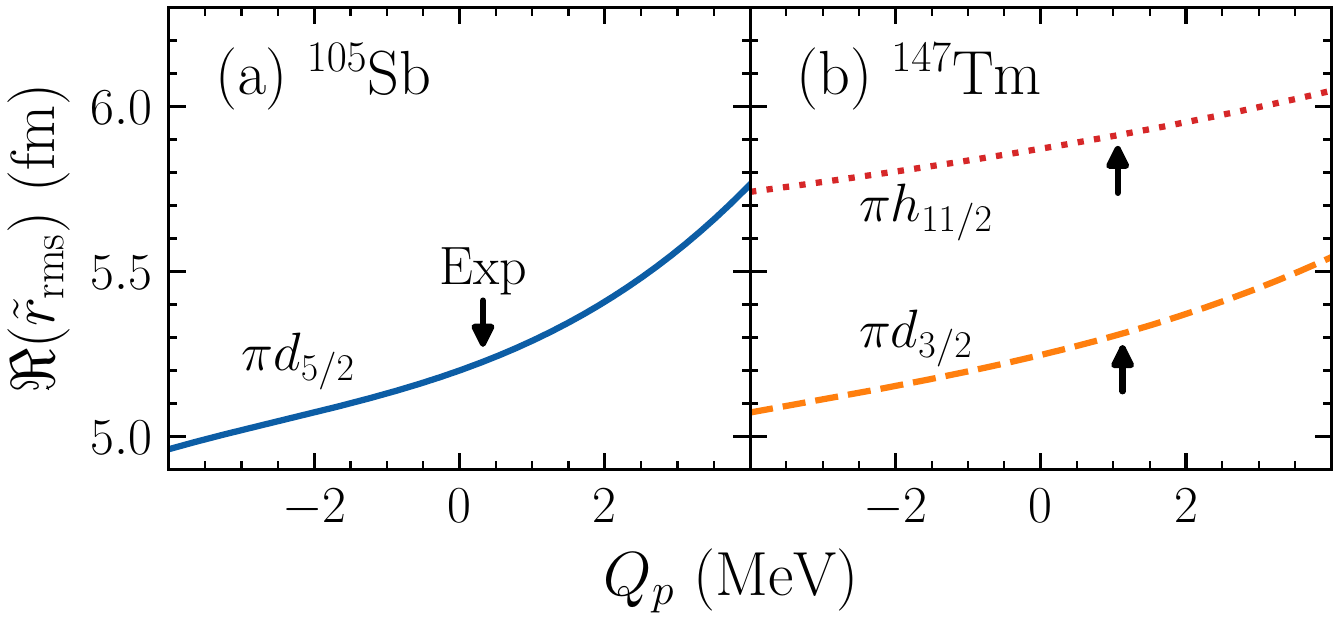}
\caption{\label{fig:longlived_system}
Real part of the valence-proton complex rms radius, $\Re(\tilde{r}_{\rm rms})$, as a function of the proton decay energy $Q_p$ for (a) the $d_{5/2}$ orbital in $^{105}\mathrm{Sb}$ and (b) the $d_{3/2}$ and $h_{11/2}$ orbitals in $^{147}\mathrm{Tm}$. Arrows mark experimental $Q_p$ values.
}
\end{figure}

\textit{Long-lived proton emitters}---To connect with expected measurements, we extend our analysis to two known long-lived proton emitters: $^{105}\mathrm{Sb}$ with $T_{1/2}=1.12$\,s \cite{Lipoglavsek2002}  and $^{147}\mathrm{Tm}$ with $T_{1/2}=0.58$\,s
\cite{Klepper1982}. For these nuclei, lifetimes exceed the timescale of the nuclear Hamiltonian ($\sim10^{-22}$\,s) \cite{Nazarewicz2001} by over 20 orders of magnitude, making time-dependent computational propagation practically impossible. Consequently, the complex-radius formulation provides a well-defined operational procedure for computing charge radii. As the resonance narrows, the plateau in Fig.~\ref{fig:t_r_cr_15F} persists longer. Figure~\ref{fig:longlived_system} shows that these long-lived systems still exhibit an enhancement of the (complex) radius with $Q_p$. It is to be noted that before comparing with experiment, the total rms charge radius requires combining the daughter-core radius with $\tilde r_{\rm rms}$ (computed in this work), including the usual mass weighting, center-of-mass, and also nucleonic and relativistic corrections \cite{Reinhard2021}.

\textit{Summary}---A complex rms radius of the resonant state is evaluated for the $d_{5/2}$ state of proton emitter \(^{15}\mathrm{F}\) using the exterior complex scaling. This quantity remains finite for unbound states, with its imaginary part being interpreted as an uncertainty associated with the decay width. Across the decay threshold, the valence-proton radius in $^{15}\mathrm{F}$ displays a nonmonotonic behavior arising from the competition between an outward shift of the density and the barrier-driven depletion of the interior. We further predict a halolike enhancement in known proton emitters $^{105}\mathrm{Sb}$ and $^{147}\mathrm{Tm}$, manifested as a smooth increase of the charge radius above the decay threshold. A complementary real-time analysis further demonstrates that, during the early stage of decay, the standard rms radius remains essentially constant and coincides with the complex rms radius, providing a practical route for computing charge radii in long-lived proton emitters. 

\textit{Acknowledgments}---This material is based upon work supported by the National Key Research and Development Program (MOST 2023YFA1606404 and MOST 2022YFA1602303); the National Natural Science Foundation of China under Contracts No.\,12347106 and No.\,12547102; and by the U.S. Department of Energy under Award No. DE-SC0013365 (Office of Science, Office of Nuclear Physics) and No. DE-SC0023175 (Office of Science, NUCLEI SciDAC-5 Collaboration).

\textit{Data availability}---The data that support the findings of this article are not publicly available.
The data are available from the authors upon reasonable
request.


%

\clearpage
\onecolumngrid
\begin{center}
\textbf{\large Supplemental Material for ``Nuclear Radii of Proton-Unbound Systems''}

\end{center}

\begin{CJK*}{UTF8}{gbsn}


\begin{abstract}
\bigskip
\noindent
This supplemental material contains:
\begin{itemize}
\item
Supplemental discussions
\item
Supplemental table
\item
Supplemental figures
\item
Supplemental videos
\end{itemize}
\end{abstract}

\maketitle
\end{CJK*}

\clearpage

\onecolumngrid
\setcounter{figure}{0}
\setcounter{table}{0}
\renewcommand{\thefigure}{S\arabic{figure}}
\renewcommand{\thetable}{S\arabic{table}}
\renewcommand{\theequation}{S\arabic{equation}}
\renewcommand{\theHfigure}{Extended Data Fig. \thefigure}
\renewcommand{\theHtable}{Extended Data Table \thetable}

\newpage
\section{SUPPLEMENTAL DISCUSSIONS}

In this section, we provide an approximate estimation of the time-dependent evolution of the rms radius \(r_{\rm rms}\).
According to the Ehrenfest theorem, the time derivative of \(r_{\rm rms}^2\) can be expressed as
\begin{equation}
\label{Ehrenfest}
\frac{d}{dt} r_{\rm rms}^2 
\equiv \frac{d}{dt}\langle r^2 \rangle 
= \frac{1}{i\hbar}\langle[\vec{r}^2, H]\rangle.
\end{equation}
Since \(\vec{r}^2\) commutes with the potential \(V\), and
\([\vec{r}^2, \vec{p}^2] = 2i\hbar\,(\vec{r}\!\cdot\!\vec{p} + \vec{p}\!\cdot\!\vec{r})\),
we obtain
\begin{equation}
\label{First_derivative}
\frac{d}{dt} r_{\rm rms}^2 
= \frac{1}{m}\,\langle \vec{r}\!\cdot\!\vec{p} + \vec{p}\!\cdot\!\vec{r} \rangle.
\end{equation}
At \(t=0\), the decay rate of the resonance vanishes~\cite{Fonda1978,Deutsch2018,Gorin2006,Volya2020},
so the first derivative of \(r_{\rm rms}^2\) is zero, and the second derivative can be evaluated as
\begin{equation}
\label{Second_derivative}
\frac{m}{2}\,\frac{d^{2}}{dt^{2}} r_{\rm rms}^2 
= 2\langle T\rangle - \langle \vec{r}\!\cdot\!\vec{\nabla} V \rangle,
\end{equation}
where \(T = \vec{p}^2/(2m)\) is the kinetic energy operator.

\begin{figure}[htbp]
\includegraphics[width=0.8\linewidth]{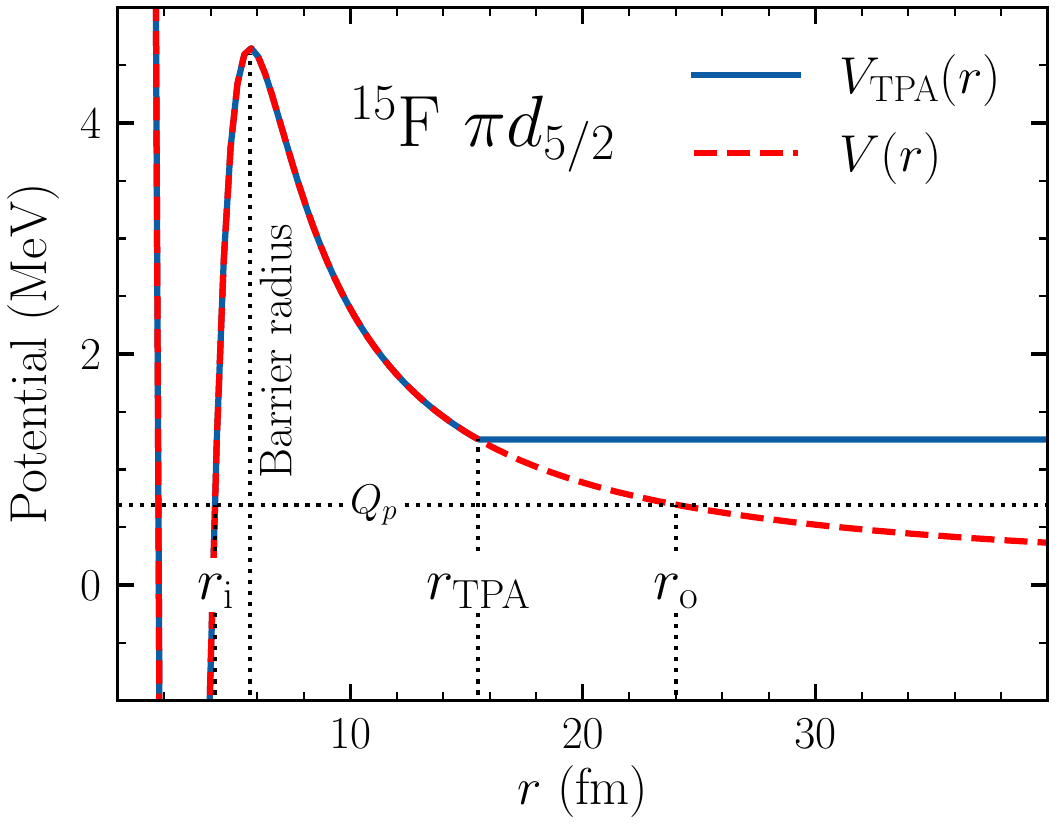}
\caption{\label{fig:TPA} The TPA potential $V_{\rm {TPA}}(r)$ \cite{Gurvitz2004} (solid line) and the potential $V(r)$ (red dashed line) as functions of $r$. The inner turning point $r_{\rm i}$ and outer turning point $r_{\rm o}$ are marked for $Q = 0.69$ MeV. The barrier radius and $r_{\rm TPA}=15.5$ fm are indicated.}
\end{figure}

The initial wave function \(\Phi(t=0)\) is prepared using the Two Potential Approach (TPA) \cite{Gurvitz2004},
in which the system is confined by an auxiliary potential \(V_{\rm TPA}\) shown in Fig.~\ref{fig:TPA}
that coincides with the physical potential \(V\) inside a chosen matching radius \(r_{\rm TPA}\),
but is constant in the outer region to produce a quasi-bound state.
According to the virial theorem, one has
\begin{equation}
\begin{aligned}
2\langle T\rangle_{t=0} &= \langle \vec{r}\!\cdot\!\vec{\nabla} V_{\rm TPA} \rangle_{t=0}, \\
\big[\,2\langle T\rangle - \langle \vec{r}\!\cdot\!\vec{\nabla} V \rangle\,\big]_{t=0} 
&= \langle \vec{r}\!\cdot\!\vec{\nabla} (V_{\rm TPA} - V) \rangle_{t=0}.
\end{aligned}
\end{equation}
Thus, Eq.~(\ref{Second_derivative}) can be approximated at \(t=0\) as
\begin{equation}
\label{Second_derivative_approximation}
\frac{m}{2}\,\frac{d^{2}}{dt^{2}} r_{\rm rms}^2 
\simeq \langle \vec{r}\!\cdot\!\vec{\nabla} (V_{\rm TPA} - V) \rangle_{t=0}.
\end{equation}
Integrating twice with respect to time gives the approximate expression
\begin{equation}
\label{r_rms_TD}
r_{\rm rms}^2(t) \simeq r_{\rm rms}^2(0) + b_{\rm TPA}\,t^2,
\end{equation}
where
\begin{equation}
\label{r_rms_TD_b}
b_{\rm TPA} = \frac{1}{m}{\langle \vec{r}\!\cdot\!\vec{\nabla} (V_{\rm TPA} - V) \rangle_{t=0}}.
\end{equation}

\begin{figure}[htbp]
\includegraphics[width=0.8\linewidth]{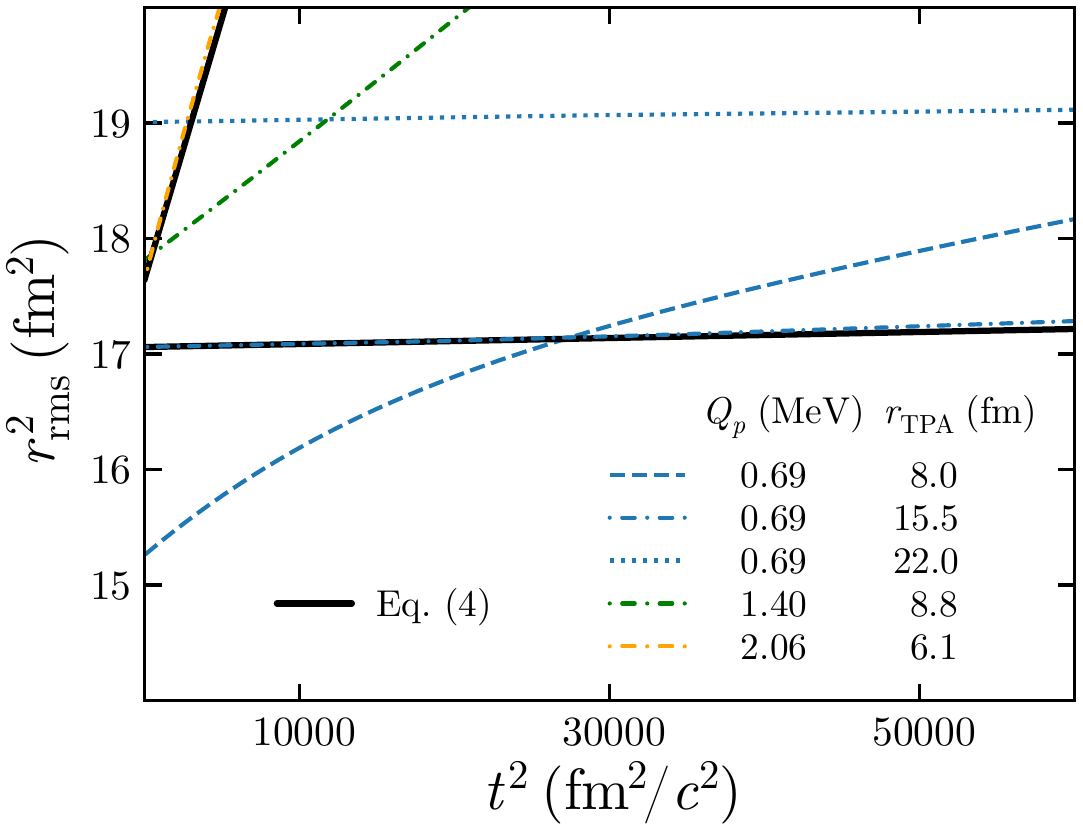}
\caption{
\label{fig:r2-t2} Calculated rms radius  \(r^2_{\rm rms}\) as a function of $t^2$. The blue lines display the results for $Q_p=0.69$ MeV with different $r_{\rm TPA}$ values: 8.0 fm (dashed line), 15.5 fm (dash-dotted line), and 22.0 fm (dotted line). The green dash-dotted line shows the result for $Q_p=1.40$ MeV with $r_{\rm TPA}=8.8$ fm. The orange dash-dotted line shows to the result for $Q_p=2.06$ MeV with $r_{\rm TPA}=6.1$ fm. The black solid line shows the analytic result ~(\ref{r_rms_TD}).}
\end{figure}

As \(b_{\rm TPA}\) depends only on the asymptotic form of the wave function at \(r>r_{\rm TPA}\), and the integrand $ \vec{r}\!\cdot\!\vec{\nabla} (V_{\rm TPA} - V) $ decays as $r^{-1}$, \(b_{\rm TPA}\)  is exceedingly small for narrow resonances. For instance, for \(^{15}\mathrm{F}\), we obtain \(b_{\rm TPA} = 2.61\times10^{-6}\,c^2\) for \(Q_p = 0.69\)~MeV and  \(r_{\rm TPA} = 15.5\)~fm, and \(b_{\rm TPA} = 4.53\times10^{-4}\,c^2\) for the \(Q_p = 2.06\)~MeV, \(r_{\rm TPA} = 6.1\)~fm, in excellent agreement with the numerical time evolution shown in Fig.~\ref{fig:r2-t2}. 

During the initial stage of the decaying process, when \(b_{\rm TPA} t^2 \ll r_{\rm rms}^2(0)\), Eq.~(\ref{r_rms_TD}) further implies
\begin{equation}
\label{r_rms_TD_expand}
r_{\rm rms}(t) \simeq \sqrt{r_{\rm rms}^2(0)+b_{\rm TPA}t^2}
\simeq r_{\rm rms}(0) + \frac{1}{2}\frac{b_{\rm TPA}}{r_{\rm rms}(0)}\,t^2,
\end{equation}
so that \(b_{\rm TPA}/r_{\rm rms}(0)\) plays the role of an effective acceleration in a ``classical-motion'' picture.

\newpage

\section{SUPPLEMENTAL TABLE}

\begin{table*}[htbp]
\caption{Dependence of the complex radius \(\tilde{r}_{\rm rms}\) of the proton $d_{5/2}$ state in  \(^{15}\mathrm{F}\) at different values of $Q_p$ on the exterior complex-scaling parameters: radius \(R_{0}\) and rotation angle \(\theta\). The residual variations reflect finite-mesh and finite-range effects and can be reduced by increasing the number of integration points and extending the integration domain.}
\label{benchmark}
\begin{threeparttable}
\begin{tabular*}{\columnwidth}{@{\extracolsep\fill}ccccc@{\extracolsep\fill}}
\toprule\toprule
$Q_p$ (MeV) & $\Gamma$ (MeV) & \multicolumn{3}{c}{$ \tilde{r}_{\rm rms}$ (fm)} \\
\cmidrule(lr){3-5} 
& & $R_0=20$ fm, $\theta=\pi/6$ & $R_0=20$ fm, $\theta=\pi/4$ & $R_0=30$ fm, $\theta=\pi/4$ \\	
\midrule
$0.69$ & $8.40\times 10^{-4}$ & $4.136+0.136i$ & $4.136+0.136i$ & $4.136+0.136i$ \\
$1.40$ & $2.48\times 10^{-2}$ & $4.251+0.503i$ & $4.251+0.503i$ & $4.247+0.504i$ \\
$2.06$ & $1.12\times 10^{-1}$ & $4.203+0.784i$ & $4.201+0.785i$ & $4.212+0.782i$ \\
\bottomrule\bottomrule
\end{tabular*}
\end{threeparttable}
\end{table*}

\newpage

\section{SUPPLEMENTAL FIGURES}

\newpage
\begin{figure}[htbp]
\includegraphics[width=0.8\linewidth]{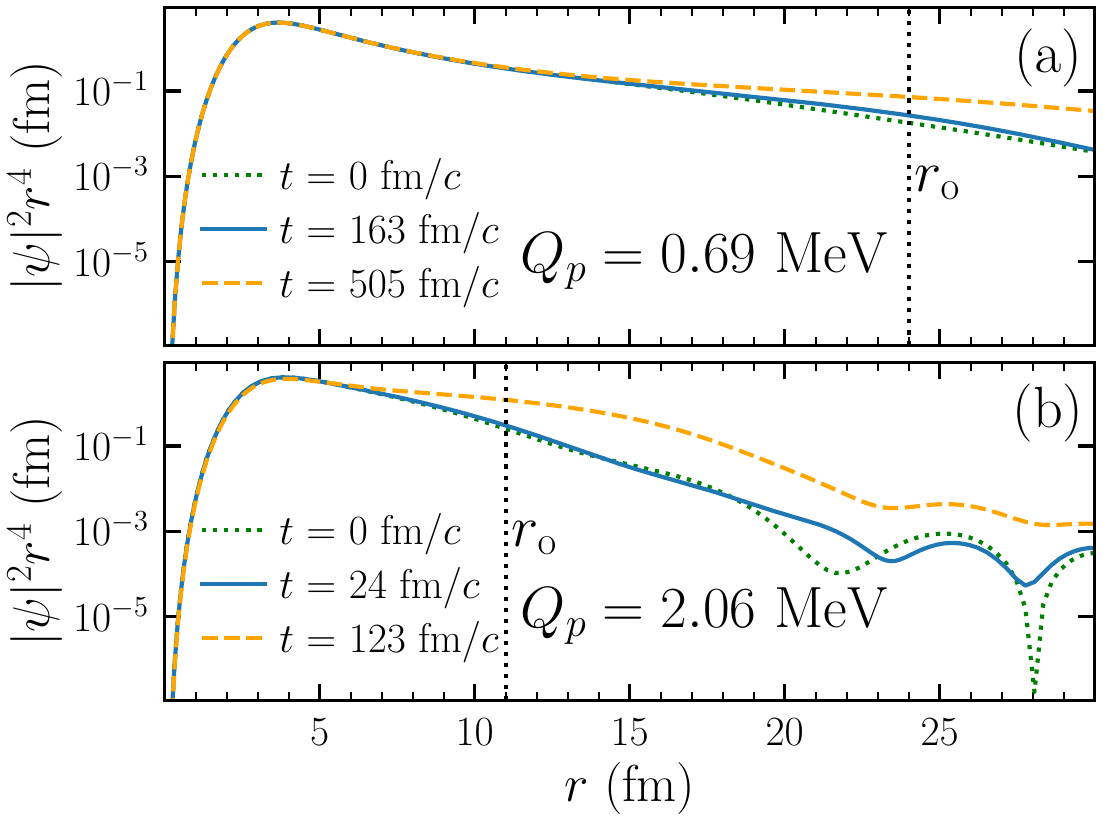}
\caption{
\label{fig:WF_diff_t_alternative} Time evolution of the integrand \( |\psi|^{2} r^{4} \) for the \(5/2^{+}\) resonance of \(^{15}\mathrm{F}\) with (a) \(Q = 0.69\)~MeV and (b) \(Q = 2.06\)~MeV   at $t=0$ (green dotted line), an early time plateau  termination  (blue solid line), and the time at which \(r_{\rm rms}\) start departing from \(\Re(\tilde{r}_{\rm rms})\) within the uncertainty 
\(\Im(\tilde{r}_{\rm rms})\) (orange dashed line). Outer turning point $r_{\rm o}$ is marked.}
\end{figure}

\newpage
\begin{figure}[htbp]
\includegraphics[width=1\linewidth]{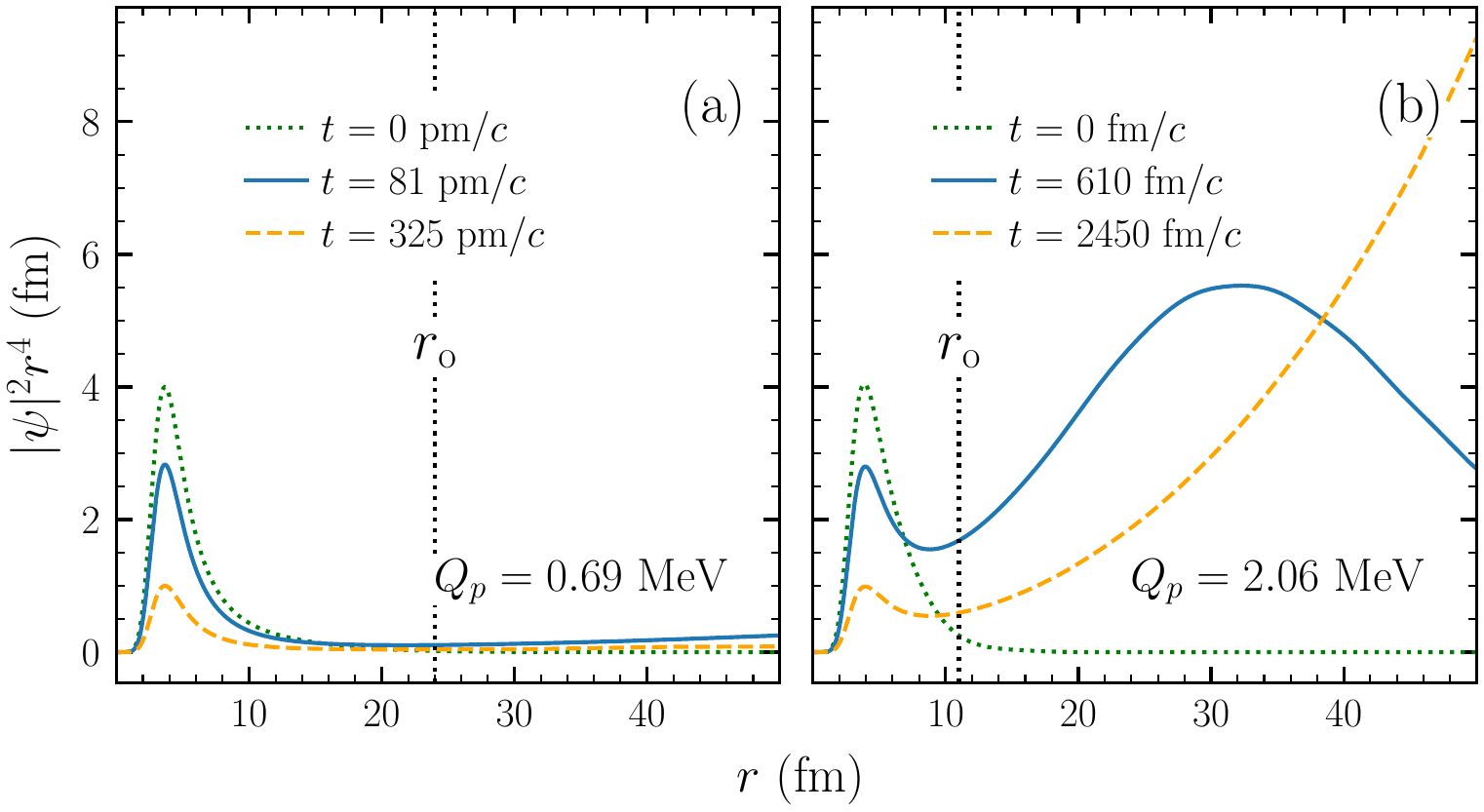}
\caption{
\label{fig:WF_diff_long_time} Time evolution of the integrand \( |\psi|^{2} r^{4} \) for (a) the \(5/2^{+}\) resonance of \(^{15}\mathrm{F}\) with \(Q = 0.69\)~MeV and (b) \(Q = 2.06\)~MeV (b) at the initial time (green dotted line), at half of the half-life (blue solid line), and at two half-lives (orange dashed line). Outer turning point $r_{\rm o}$ is marked.}
\end{figure}

\newpage

\newpage
\begin{figure}[htbp]
\includegraphics[width=1\linewidth]{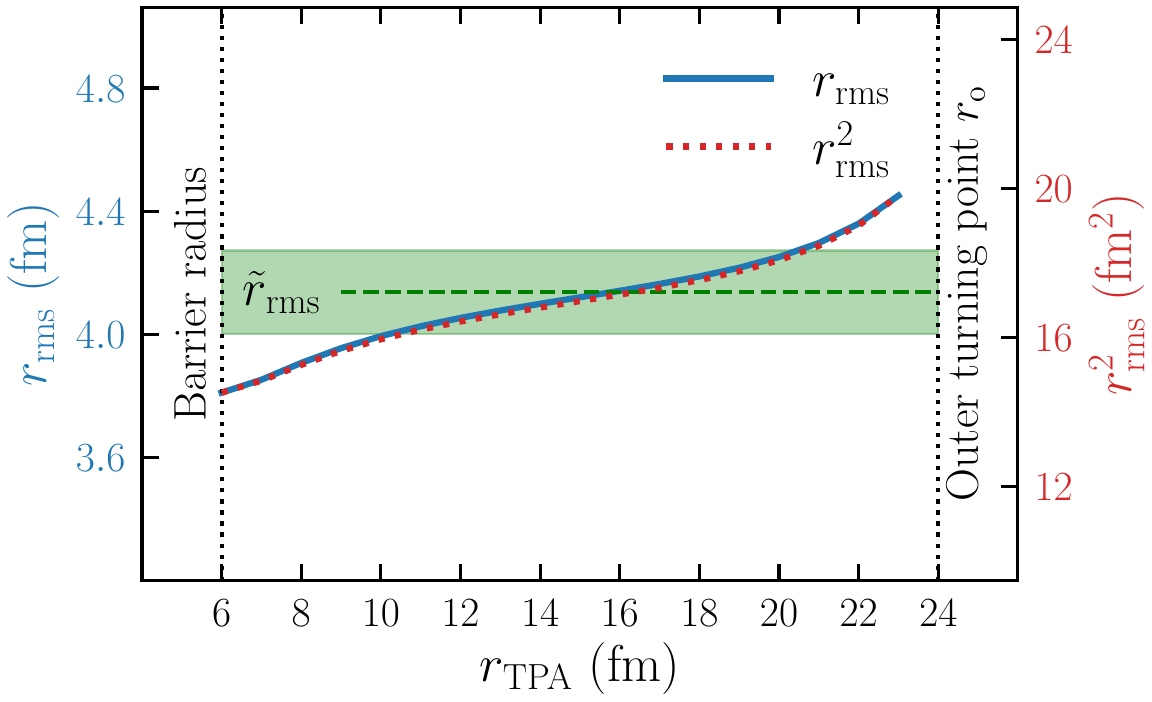}
\caption{\label{fig:r_rms_vs_r_TPA} Calculated rms radius \(r_{\rm rms}\) (blue solid line) and its square \(r^2_{\rm rms}\) (red dotted line) for \(^{15}\mathrm{F}\) at
\(Q_p = 0.69\) MeV and \(t = 0\)
as a function of \(r_{\rm TPA}\). The range of \(r_{\rm TPA}\) spans from the barrier radius (\(\sim 6\) fm) to the value slightly below the outer turning point at \(r_{\rm o} \approx 24\) fm. The complex radius is also shown (green band).}
\end{figure}

\newpage
\begin{figure}[htbp]
\includegraphics[width=0.8\linewidth]{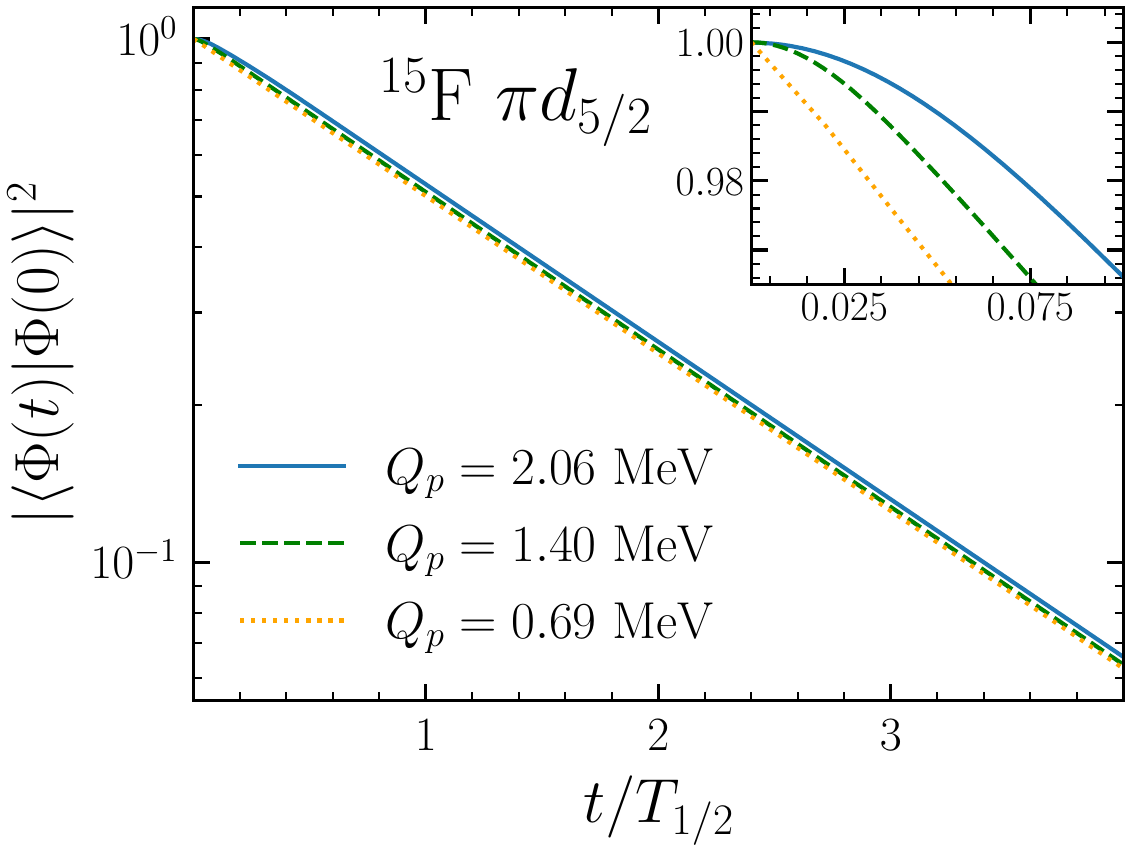}
\caption{\label{fig:SP} Survival probability $|\langle \Phi(t)|\Phi(0)\rangle|^2$ as a function of time (relative to $T_{1/2}$) for $^{15}$F at various decay energies: $Q_p=2.06$ MeV (blue solid line), $Q_p=1.40$ MeV (green dashed line) and $Q_p=0.69$ MeV (orange dotted line).}
\end{figure}

\newpage

\section{SUPPLEMENTAL VIDEOS}

These supplemental videos illustrate the time evolution of the radial probability distribution and its cumulative contribution to the rms radius of the unbound nucleus \(^{15}\mathrm{F}\). The visualizations demonstrate the sensitivity of the wave-function dynamics to the truncation of the initial spatial configuration and to the decay energy \(Q_p\).

\begin{itemize}
\item
\href{https://www.dropbox.com/scl/fi/6rq0h5d4besb7wa77w6pp/15F_snr2_50d_box50.mov?rlkey=hawa3jlpxdz88p1drb8owypzg&st=es9kqw8s&dl=0}{Supplementary video 1}: Time evolution of the integrand \( |\psi|^{2} r^{4} \), which determines the rms radius \(r_{\mathrm{rms}}\), for the \(5/2^{+}\) resonance of \(^{15}\mathrm{F}\) with \(Q_p = 0.69\)~MeV. 
The animation compares different spatial truncations of the initial wave function (\(r < R_{\mathrm{cut}}\)).

\item
\href{https://www.dropbox.com/scl/fi/1pcurrzhi290bomxeyyj6/15F_snr2_46d_box50.mov?rlkey=gf7rclpv4nz0g5o4132dqaqej&st=6wq6lk2b&dl=0}{Supplementary video 2}: Same as  Supplementary video 1, but for a higher decay energy \(Q_p = 2.06\)~MeV.

\item
\href{https://www.dropbox.com/scl/fi/zlsc1qmpvhx0ktin8706d/15F_int_snr2_50d_box50.mov?rlkey=gzxxsxey7obq8jzsv4sifbm43&dl=0}{Supplementary video 3}: Evolution of the cumulative integral $
\int_{0}^{r} |\psi(r')|^{2} r'^{4}\,dr'
$ for the \(5/2^{+}\) state of \(^{15}\mathrm{F}\) at \(Q_p = 0.69\)~MeV.

\end{itemize}

\end{document}